%% file: CDC-Backdoor.tex
\documentclass[letterpaper,10pt,conference]{ieeeconf}
\usepackage{cite}
\usepackage{graphicx}
\pdfoutput=1
\usepackage{subcaption}
\usepackage[cmex10]{amsmath}
\usepackage{algorithm}
\usepackage{algpseudocode}
\usepackage{xcolor}
\usepackage{array}
\usepackage{url}
\usepackage{bbm}
\usepackage{todonotes}
\usepackage{color,soul}
\usepackage{epsfig}
\usepackage{amsfonts,amssymb,multirow,bigstrut,booktabs,ctable,latexsym}
\usepackage{mathtools}
\usepackage[mathscr]{eucal}
\usepackage{verbatim}

\usepackage{amsthm}
\usepackage{soul}
\usepackage[normalem]{ulem}

\newcommand{\ie}{\text{i.e., }}

\usepackage{algpseudocode}
\IEEEoverridecommandlockouts

\title{Trojan Horse Training for Breaking Defenses against\\ Backdoor Attacks in Deep Learning }
	
	\author{Arezoo Rajabi$^{1}$, Bhaskar Ramasubramanian$^{2}$, Radha Poovendran$^{1}$%
		\thanks{$^{1}$Network Security Lab, Department of Electrical and Computer Engineering, 
			University of Washington, Seattle, WA 98195, USA. \newline
			{\tt\small \{rajabia, rp3\}@uw.edu}}
\thanks{$^2$Electrical and Computer Engineering, Western Washington University, Bellingham, WA 98225, USA.
			{\tt\{ramasub\}@wwu.edu}}
	}
\begin{document}	
	\maketitle
	
	
\begin{abstract}
Machine learning (ML) models that use deep neural networks are vulnerable to backdoor
attacks. Such  attacks involve the insertion of a (hidden) trigger by an adversary. As a
consequence, any input that contains the trigger will cause the neural network to misclassify
the input to a (single) target class, while classifying other inputs without a trigger correctly. ML
models that contain a backdoor are called \emph{Trojaned models}. Backdoors can have severe
consequences in safety-critical cyber and cyber physical systems when only the outputs of the
model are available. Defense mechanisms have been developed and illustrated to be able to
distinguish between outputs from a Trojaned model and a non-Trojaned model in the case of a
single-target backdoor attack with accuracy $>96\%$.

Understanding the limitations of a defense mechanism requires the construction of examples
where the mechanism fails. Current single-target backdoor attacks require one trigger per
target class. We introduce a new, more general attack that will enable a single trigger to result
in misclassification to more than one target class. Such a misclassification will depend on the
true (actual) class that the input belongs to. We term this category of attacks \emph{multi-target
backdoor attacks}. We demonstrate that a Trojaned model with either a single-target or multi-
target trigger can be trained so that the accuracy of a defense mechanism that seeks to
distinguish between outputs coming from a Trojaned and a non-Trojaned model will be
reduced. Our approach uses the non-Trojaned model as a ‘teacher’ for the Trojaned model and
solves a min-max optimization problem between the Trojaned model and defense mechanism. 
Empirical evaluations demonstrate that our training procedure reduces the accuracy of a state-of-the-art defense mechanism from $>96\%$ to $0\%$. 
We also discuss possible approaches to improve
defense mechanisms to ensure resilience to backdoor attacks for a broader category of ML models. 
\end{abstract}

\input{Intorduction}
\input{Preliminary}
\input{relatedwork}
\input{ProposedMethod}
\input{Experiments}
\input{discussion}
\input{conclusion.tex}
\bibliographystyle{IEEEtran}
\bibliography{MainRef}
\end{document}

%% file: Intorduction.tex
\section{Introduction}
%
%
The recent advances in cost-effective storage and computing has resulted in the wide use of deep neural networks (DNN) across multiple data-intensive applications such as face-recognition~\cite{taigman2014deepface}, mobile networks~\cite{zhang2019deep}, computer games~\cite{mnih2015human}, and healthcare~\cite{esteva2019guide}. 
The large amounts of data and extensive computing resources required to train these deep networks has made online machine learning (ML) platforms~\cite{AWS, BigML, Caffe} increasingly popular. 
However, these platforms only provide access to input-output information from the models, and not parameters of the models themselves. 
This is termed \emph{black-box access}. 
Recent research~\cite{gu2019badnets} has demonstrated that online ML models can be trained in a manner so that the presence of a specific perturbation, called a \emph{trigger}, in the input will result in an output that is different from the correct or desired output. 
At the same time, outputs of the model for clean inputs- i.e., inputs without trigger- is not affected. 
This can result in severe consequences when such platforms are used in safety-critical cyber and cyber-physical systems~\cite{ullah2019cyber}. 
The insertion of a trigger into inputs to a model is called a \emph{backdoor attack} and the model that misclassifies such inputs is termed \emph{Trojaned}. 
 
Backdoor attacks can have severe implications in safety-critical cyber and cyber-physical systems where only the outputs of a model are available. 
For example, autonomous navigation systems that depend on DNNs for decision making using reinforcement learning models have been shown to be vulnerable to backdoor attacks~\cite{panagiota2020trojdrl}. 
DNN models have also been used for traffic sign detection, and these models can be trained to identify the signs correctly, but results in an incorrect output when the sign has a trigger~\cite{gu2019badnets} (e.g., a `stop' sign with a small sticker on it is identified as a `speed-limit' sign). 
Such threats necessitate the development of defense mechanisms. 
However, most defenses assume that hyper-parameters of the model are available~\cite{liu2018fine,yoshida2020disabling,li2021neural,kolouri2020universal} or that a pre-processing module can be added to the target model~\cite{liu2017neural}. 
These may not be practical for applications where only outputs of the model are available and users cannot control inputs provided to the model.

A defense mechanism against backdoor attacks called meta-neural Trojan detection (MTND) was proposed in~\cite{xu2021detecting}. 
This method leverages an insight that the distribution of outputs from a Trojaned model might be different to those from a non-Trojaned model, even though both models have similar accuracies. 
MTND learns a discriminator (a classifier with two outputs, YES or NO) using outputs from a Trojaned and a non-Trojaned model as training data, in order to distinguish between the models. 
This approach was shown to identify Trojaned models with $>96\%$ accuracy, when only black-box access to them was available~\cite{xu2021detecting}. 
Despite the current success of MTND, examples can be constructed that demonstrate some of its limitations.
 
%
 \begin{figure*}
    \centering
    \includegraphics[width=0.64\textwidth]{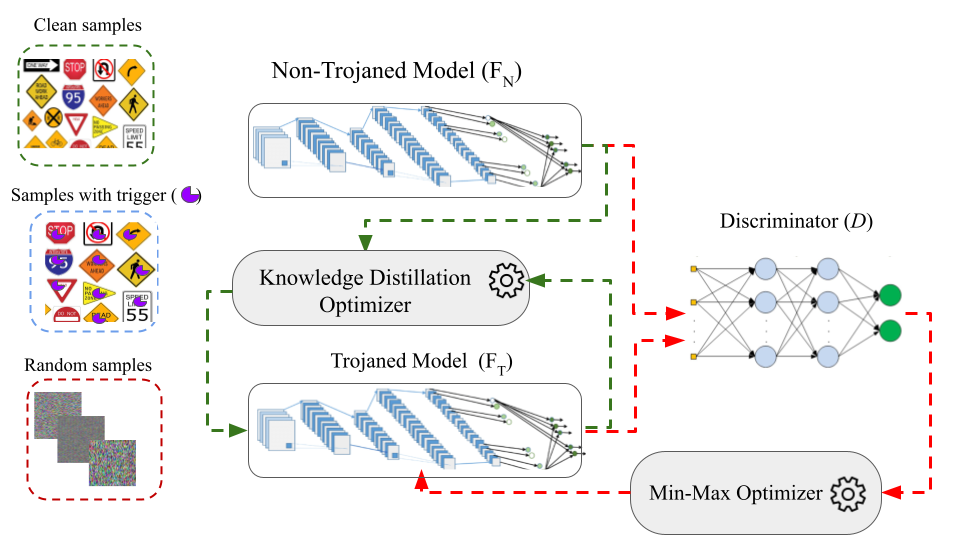}
    \caption{Schematic of our two-step methodology. \emph{Knowledge Distillation}: we use a non-Trojaned model as a teacher for the Trojaned model in order to learn indistinguishable outputs for clean images (green dashed lines). \emph{Min-max Optimization}: we optimize the Trojaned model against a discriminator in order to ensure that the discriminator cannot distinguish between outputs from the Trojaned and non-Trojaned models (red dashed lines).}
    \label{fig:scheme}
\end{figure*}
%
 %

In this paper, we identify a new class of backdoor attack called {\it multi-target backdoor attack}. Unlike existing single-target trigger backdoors, a trigger from a multi-target backdoor attack can cause misclassification to different output labels depending on the true class of the input. Specifically, we demonstrate that a model can be trained so that its output is a function of  the true class of the input and the presence of a trigger in that input. 
We also propose a two-step methodology to bypass the MTND defense mechanism.  
Figure~\ref{fig:scheme} demonstrates our approach: 
(i) we use a non-Trojaned model as a \emph{teacher} for the Trojaned model (\textbf{Knowledge Distillation}), then (ii) we use min-max optimization between a discriminator and Trojaned model to ensure that the discriminator is not able to distinguish between outputs from a Trojaned and a non-Trojaned model (\textbf{Min-max Optimization}).

We make the following contributions:

\begin{itemize}
    \item We introduce a new class of \emph{multi-target backdoor attacks}. 
    Such an attack has the property that a single trigger can result in misclassification to different output labels, based on true input label.
    
   \item We design two algorithms- a training procedure that combines knowledge distillation (Algorithm 1) and min - max optimization (Algorithm 2) to reduce the accuracy of a defense mechanism designed to distinguish between Trojaned and non-Trojaned models. 
   \item We evaluate the trained Trojaned model from the previous step by examining the effect of a multi-target backdoor attack on a state-of-the-art meta-neural Trojan defense (MTND). Our empirical evaluations demonstrate that our training procedure is able to bypass the MTND defense $100\%$ of the time.
    
   
   
\end{itemize}

The remainder of this paper is organized as follows: 
Section \ref{sec:preliminaries} presents a tutorial introduction to DNNs with backdoors and describes our system model. 
An overview of related literature on backdoor attacks in deep learning and state-of-the-art defense mechanisms is provided in Section \ref{sec:relatedwork}. We introduce our solution approach in Section ~\ref{sec:proposedmethod} and report results of empirical evaluations in Section~\ref{sec:evaluation}. 
Section \ref{sec:discussion} discusses methods to extend our solution to a broader class of problems, and Section \ref{sec:conclusion} concludes the paper.

%% file: Preliminary.tex
\section{Preliminaries}~\label{sec:preliminaries}
This section provides a brief introduction to classification using deep neural networks, and single-target backdoor attacks using a \emph{set of poisoned inputs}. Finally, we introduce the system model that we use for our algorithms.

\subsection{Deep Neural Networks}
Deep Neural Network (DNN) classifiers are trained to predict the  most relevant class among $C$ possible classes for a given input.  
The output of a DNN is called a \emph{logit}, which gives a weight to each class, $z:=[z^1,\cdots, z^C]$. The output of the model is fed to the softmax function to generate a probability vector where each element $i$ is the conditional probability of class $i$ for a given input $x$. The softmax function is defined as:
\begin{equation}\label{eq:softmax}
     p(z^i,T) = \frac{\exp{z^i/T}}{\sum_j^C \exp{z^j/T}}, 
\end{equation}
where $T$ is a temperature parameter (typically $=1$). 
A DNN classifier is a function $z:=F(x;\theta)$, where $x\in[0,1]^{d}$ is an input and $\theta$ represents hyperparameters of the DNN. 
We will write $p(z,T)$ to denote the probability vector determined through the softmax function. 
In order to train the DNN (i.e., determine values of $\theta$), we minimize the difference between the output of softmax function $p(F(x_k;\theta),T=1)$, and the true class of the input, $y^*_k$ for a sample $x_k$. 
This is quantified by a loss function $\mathcal{L}(p, y^*)$, and 
%
parameters $\theta$ are iteratively updated using stochastic gradient descent as: 
\begin{equation}\label{eq:LCE}
    {\theta}^{t+1}\gets \theta^t -\alpha \frac{1}{|\mathcal{D}|}\sum_k \frac{\partial}{\partial \theta}  \mathcal{L}(p(F(x_k;\theta)),y_k^*)
\end{equation}
 where $\mathcal{D}$, $F$, $\alpha$ and $\mathcal{L}$ are training set, DNN's function with hyper-parameter $\theta$, a positive coefficient  and loss function respectively.

One way of introducing a backdoor into the model is through poisoning the training set with a set of inputs stamped with a pre-defined trigger and labeled with the desired output~\cite{liu2020reflection,li2020rethinking}. 
The trigger has a single-target \ie any input with trigger causes the model to return a specific output. 
In order for a model to return multiple target classes, we will require one trigger per target class to be inserted into the input. 
%
Let $\mathcal{D}=\{(x_1,y_1),(x_2,y_2),\cdots, (x_N,y_N) \}$ be the original training set (a set of clean samples) and $\mathcal{D'}=\{ (x_{1}',y^d), (x_{2}',y^d),\cdots, (x_{n}',y^d)\}$ ($n \ll N$) be a set of  perturbed samples. 
Suppose each sample in  $\mathcal{D'}$ is perturbed using a pre-defined trigger as: 
 \begin{align*}
     x_{ij}'=  m_{ij}*\Delta + (1-m_{ij}) x_{ij} \:\:\ie\:\: i\in [1,W],\:\: j\in [1,H]
 \end{align*}
 where $\Delta$ is the perturbation that we term a Trojan trigger and $m$ is a mask that indicates the location where the perturbation is applied. 
 In this paper, we assume that each sample is an image of resolution $W\times H$. 
The trained Trojaned  model on both clean and poisoned datasets would return a desired output in the presence of a specific trigger in the input while keeping the accuracy unchanged for clean samples. 
 

\subsection{Our System Model}
In this paper, we assume that the Trojaned model is trained by an adversary who does not share the hyper-parameters of her model. 
The Trojaned model can be shared through an ML platform or can be a built-in model in a smart device.  Therefore, only the outputs of the model are available to users/defenders for any given  input. This is termed \emph{black-box access}.  
The defender aims to learn a discriminator (a classifier with two classes of YES/NO) to determine whether a model is Trojaned or not. The defender can learn several non-Trojaned and Trojaned models locally and use their outputs to train the discriminator (See Figure~\ref{fig:scheme}).  
We also assume the defender and adversary have access to the same training sets to train their local models. 
Given an arbitrary set of inputs that is provided to both a Trojaned and non-Trojaned model, the discriminator uses the outputs from these two models to learn a (binary) classifier.  
After training, the discriminator is used to evaluate an unknown model, to determine whether it is Trojaned or not. 
Our contribution in this paper is the design of a methodology to demonstrate that such a discriminator can be fooled (i.e., cannot say whether a model is Trojaned with probability $>0.5$). 

%% file: relatedwork.tex
\section{Related Work}
\label{sec:relatedwork}
This section summarizes recent literature on backdoor attacks and defense mechanisms against such attacks. 
%

A backdoor attack results in DNN models misclassifying inputs that contain a trigger~\cite{usenix2021blind}. 
An input containing a trigger can be viewed as an adversarial example~\cite{kurakin2017adversarial}, but there are differences in the ways in which an adversarial example attack and a backdoor attack are carried out. 
Adversarial examples are generated by learning a quantum of adversarial noise that needs to be added to an input in order to cause a pre-trained DNN model to misclassify the input~\cite{moosavi2016deepfool,kurakin2017adversarial,goodfellow2015explaining}. 
Backdoor attacks, in comparison, aim to influence weights of parameters of neural networks that describe the target model during the training phase either through poisoning the training data~\cite{usenix2021blind} or poisoning the weights themselves~\cite{dumford2020backdooring,rakin2020tbt}. Consequently, any input stamped with a pre-defined trigger will be able to cause the DNN to misclassify the input. 
Moreover, the trigger for a backdoor attack might be invisible~\cite{turner2019label, li2020invisible, liao2018backdoor}
It has been demonstrated that adversarial example attacks and backdoor attacks can both be trained to be transferable~\cite{gu2019badnets,wang2020backdoor}. 
When access to a target model was not available, a black-box backdoor attack was proposed in~\cite{liu2017trojaning}. 
In this case, the adversary had to generate training samples, since the training dataset was not available. 
%

The vulnerability of DNN models to backdoor attacks have been examined in multiple applications. 
A class of semantic backdoor attacks was proposed in~\cite{bagdasaryan2020backdoor,usenix2021blind} for image and video models. 
In this case, the target label was determined based on features of the input- for e.g., inputs featuring green colored cars would result in the model misclassifying it as a bicycle. 
The authors of~\cite{usenix2021blind} and ~\cite{dai2019backdoor} designed backdoor attacks for code poisoning and natural language processing models respectively. 
Recently, an untargeted backdoor attack for deep reinforcement learning models was proposed in ~\cite{panagiota2020trojdrl}. 
Backdoor attacks have also been used to determine fidelity- specific examples include model watermarking~\cite{adi2018turning} and verifying that a request by a user to delete their data was actually carried out by a server~\cite{sommer2020towards}. 
 %

Backdoor attacks typically specify one target output class for each trigger. 
An \emph{N-to-one trigger} backdoor attack was proposed in~\cite{xue2020one}, where an adversary used a single trigger at a specific location, but with different intensities for each target. 
A procedure to train different models to output different target classes using the same trigger was demonstrated in~\cite{xiao2022multitarget}. 
The authors of~\cite{rajabi2020adversarial} introduced an input-based misclassification procedure by learning adversarial perturbations~\cite{moosavi2017universal} for pairs of target classes. 
Different from~\cite{rajabi2020adversarial} where one perturbation was required for each pair of classes, our methodology in this paper uses a single trigger that can cause misclassification to more than one target class.   
%
%

%

Defense mechanisms against backdoor attacks can focus on removing backdoors from Trojaned models~\cite{liu2018fine,yoshida2020disabling,li2021neural} or detecting/suppressing poisoned data during training~\cite{du2019robust,tran2018spectral,chen2018detecting}. 
Our focus in this paper is on a third type- defense mechanisms for pre-trained models, since we assume that the target model has already been trained, and that the user will only have `black-box' access to it. 
%

Pre-processing defense mechanisms deploy a module that will remove or reduce the impact of a trigger present in the input. 
For example, an auto-encoder was used as a pre-processor in~\cite{liu2017neural}. 
A generative adversarial network (GAN) was used to identify `influential' portions of an image/ video input in~\cite{selvaraju2017grad}. 
This approach was leveraged by~\cite{udeshi2019model} to use the dominant color of the image as a trigger-blocker. 
Style-transfer was used as a pre-processing module in~\cite{villarreal2020confoc} to mitigate the impact of trigger present in the input. 
Modifying the location of the trigger using spatial transformations was deployed as a defense mechanism in~\cite{li2020rethinking}. 
%

In contrast, post-training defense mechanisms aim to determine whether a given model is Trojaned or not, and then refuse to deploy a Tojaned model. 
The authors of~\cite{kolouri2020universal} proposed a technique based on learning adversarial perturbations~\cite{moosavi2017universal} to locate a trigger using an insight that triggers constrain the magnitude of the learned perturbation. 
Thus, the learning process would identify a model as Trojaned if the learned perturbation was below a threshold. 
An outlier detector was used in~\cite{huang2019neuroninspect} to explain outputs of a model and features extracted using a saliency map were used to identify if the model was Trojaned. 
A defense mechanism against backdoor attacks when working with limited amounts of data was proposed in~\cite{wang2020practical}.  
%

All the approaches described above require access to hyperparameters of the model. 
There is a relatively smaller body of work focused on designing defenses in the absence of such access. 
A mechanism called DeepInspect was proposed in~\cite{chen2019deepinspect}, which learned the probability distribution of triggers using a generative model. 
DeepImpact assumed that a trigger had fixed patterns with a constant-valued distribution. 
In comparison, we consider a trigger that can have an arbitrary location within the input, and can result in misclassification to more than one target class. 
The authors of~\cite{xu2021detecting} proposed meta neural Trojan detection (MTND). 
MTND used a discriminator which took a target model as input, and performed a binary classification on the output of the model to identify if it was Trojaned or not. 
We evaluate our methodology using MTND as a benchmark, since MTND does not make any assumptions about the trigger. 
%

%% file: ProposedMethod.tex
\section{Solution Approach} \label{sec:proposedmethod}

Backdoor attacks aim to preserve the accuracy of a model on inputs without a trigger (clean samples) while misclassifying inputs that are stamped with a trigger. 
We denote the (Trojan) trigger by $\Delta$. 
Different from existing backdoor attacks that result in the model producing a single, unique target class for inputs with a trigger, we propose a new class of \emph{multi-target backdoor attacks}. 
A multi-target backdoor attack can result in a model producing a different output based on the true class of the input that contains a trigger. 
Consequently, an adversary can trigger a desired output by selecting a sample from the corresponding source class. 
%

In order to train a multi-target backdoor, an adversary poisons the training input set with a new set of samples perturbed with the trigger, and labeled using a map-function: 
\begin{align*}
    \mathcal{D'}&=\{(x_{i_1}+m\Delta,g(y_{i_1}^*)), (x_{i_2}+m\Delta,g(y_{i_2}^*)),\\&\qquad \cdots, (x_{i_n}+m\Delta,g(y_{i_n}^*))\}, 
\end{align*}
where $\Delta, m, g(\cdot)$ are the trigger, a mask which denotes where the trigger is deployed, and a function that maps a source class to target class (that is, $g(i) = j, i, j \in \{1,2,\dots C\}, i \neq j$) respectively. 
However, a backdoor attack can result in output distributions (i.e., probabilities that the output belongs to a specific class) from a Trojaned model being different to that from a non-Trojaned model, even though both models will have similar accuracy on clean samples. 
%

\begin{figure}
    \centering
    \includegraphics[width = 0.5 \textwidth]{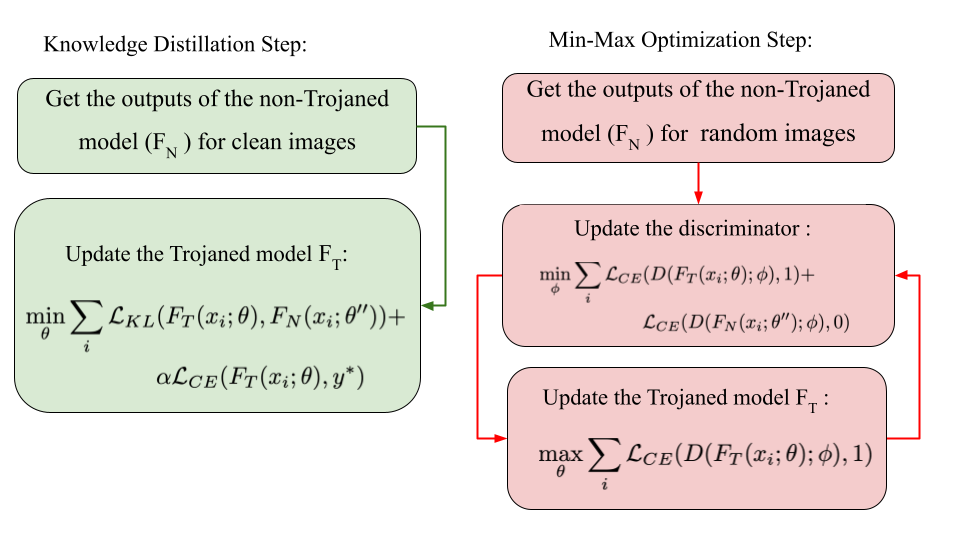}
    \caption{Two steps of our learning procedure. In the first step (left chart), we use a non-Trojaned model as a teacher for the Trojaned model. Then in the second step (right chart), we update the weights of model to fool the discriminator.}
    \label{fig:LearningScheme}
\end{figure}

Our objective is to demonstrate the limitations of defense mechanisms that seek to distinguish between outputs from a Trojaned model and a non-Trojaned model can be bypassed. 
To this end, we seek to learn a Trojaned model that has an output distribution on clean samples which is similar to that from a non-Trojaned model. 
The two-stage setup that we use is shown in Figure~\ref{fig:LearningScheme}. 
We first learn such a Trojaned model using a non-Trojaned model as a teacher. 
Then, we maximize the indistinguishability between outputs of the two models by solving a min-max optimization problem. 
The remainder of this section explains these steps in detail. 
%

\subsection{Using a non-Trojaned Model as Teacher}

The process of transferring knowledge from a `teacher' model that has high accuracy to a smaller `student' model is called \emph{knowledge distillation}~\cite{hinton2015distilling}. 
Unlike `hard' labels which assign a sample to exactly one class with probability $1$, `soft' labels assign a probability distribution over output classes. 
The non-zero probabilities provide information about inter-class similarities. 
Knowledge distillation improves accuracy of the student model using `soft' labels provided by the teacher. 
In order to ensure similarities in behaviors of a Trojaned model and a non-Trojaned model, we use the non-Trojaned as a teacher for the Trojaned model, and minimize the difference between their outputs for a given input. 


Let the DNNs that comprise the Trojaned model be parameterized by $\theta$, and those comprising the non-Trojaned model be parameterized by $\theta''$. 
Assume that $z_{T} = F_{T}(.;\theta)$ and $z_N = F(.;\theta'')$ denote the logits from the two models, and $C$ be the number of output classes. 
Knowledge distillation techniques minimize the distance between the outputs of teacher and student models for clean samples. 
The authors of~\cite{hinton2015distilling} established that when using gradient-based techniques and soft labels  as an input to a loss function, the gradients were scaled by $1/T^2$. 
For e.g., when using the $L_2$-norm to measure the distance between the models' outputs, following Eqn (\ref{eq:softmax}) and assuming that logits are zero mean without loss of generality, the gradient of the loss function is:
\begin{align*}
    &\frac{\partial }{\partial z^i_{T}} 0.5(p(z_{N}, T)-p(z_{T}, T))^2 \\
    &=\frac{1}{T} \frac{e^{z^i_{T}/T}}{\sum_j e^{z^j_{T}/T}} (\frac{e^{z^i_{N}/T}}{\sum_j e^{z^j_{N}/T}} - \frac{e^{z^i_{T}/T}}{\sum_j e^{z^j_{T}/T}}) \\
   &\approx \frac{1}{T} \frac{1+z^i_{T}/T}{C+\sum_j z^j_{T}/T}(\frac{1+z^i_{N}/T}{C+\sum_j z^J_{N}/T}-\frac{1+z^i_{T}/T}{C+\sum_j z^j_{T}/T})  \\
   &\approx \frac{1}{C^2T^2}z_T^i (z_N^i-z^i_{T} )
\end{align*}
To compensate for the $1/T^2$ factor obtained in the above gradient-calculation, we will scale by $T^2$ when updating parameters $\theta$ of the DNN (see \emph{Line 12} of Algorithm 1). 
The gradients for other loss functions like the Kullback-Liebler (KL)-divergence, $\mathcal{L}_{KL}$, and cross-entropy loss, $\mathcal{L}_{CE}$ can be computed in a similar manner. 

The Kullback-Leibler divergence loss function quantifies the distance between the distributions of the outputs of two models and is defined as: 
\begin{equation}
    \begin{split}
    \mathcal{L}_{KL} (p(z_N, T), p(z_T,T))= p(z_N, T) \log \frac{p(z_N, T)}{p(z_T,T)}
    \end{split}
\end{equation}
%

Knowledge distillation techniques also train the Trojaned model on hard labels together with soft labels. 
The cross entropy loss function measures the difference between the the softmax  of the logit of the training (Trojaned) model, denoted $T=1$, and the true label of the input as follows:
\begin{equation}
\begin{split}
    \mathcal{L}_{CE} (y^*, p(z_{T},T=1))&=-\sum_j y_j^* \log p_j(z_{T},T=1)\\
    &= -\log p_i(z_{T},T=1) 
\end{split}
\end{equation}
where $y^*$ is the true class of the input and $p_i(z_T,1)$ is the $i^{th}$ element of the softmax function's output. Since a hard label ($y^*$) assigns an input to one class, all elements have value of $0$, except the $i^{th}$ element, which has a values of $1$ if the input belongs to class $i$.  Algorithm~\ref{alg:KD} explains the knowledge distillation step in detail. 

\begin{algorithm}
\caption{Knowledge Distillation}\label{alg:KD}
\begin{algorithmic}[1]
\Require $\mathcal{D}, \Delta, m, g(.), T^*, F_{N}, \alpha_1>0, \alpha_2>0$
\State $\mathcal{D'}\leftarrow \mathcal{D}$
\For {$j=1:n$}
\State $(x,y)\leftarrow \text{random-selection}(\mathcal{D})$
\State $\mathcal{D'}\leftarrow \mathcal{D'} \cup (x+m\Delta, g(y))$ 
\EndFor

\For {$i=1:itr$}
\For {$(x^k,y^k) \in \mathcal{D}$}

\State $z_{T} \gets F_{T}(x^k;\theta)$
\State $z_{N} \gets F_{N}(x^k;\theta'')$
\State $q_1 \gets p (z_{N}, T=T^*)$
\State $q_2 \gets p (z_{T}, T=T^*)$
\State  $L_1 \leftarrow L_1+ T^2   \times \frac{\partial}{\partial \theta} \mathcal{L}_{KL}(q_1,q_2)$
\EndFor
\For {$(x^k,y^k) \in \mathcal{D'}$}
\State $z^k_{T} \gets F_{T}(x^k;\theta)$
\State $L_2\leftarrow L_2+ \times \frac{\partial}{\partial \theta} \mathcal{L}_{CE} (y^k, p(z_{T}, T=1)) $
\EndFor
\State $\theta \gets \theta- \alpha_1 L_1 - \alpha_2 L_2 $
\EndFor
\end{algorithmic}
\end{algorithm}
 
\subsection{Min-Max Optimization}
We assume that a defender has access to only the outputs of a Trojaned model. 
This is termed \emph{black-box access}, and is a reasonable assumption when machine learning-enabled cyber and cyber-physical systems are deployed in the real world. 
In order to determine whether a model is Trojaned or not using only outputs of the model, the defender uses a \emph{discriminator}, $D$. 
The discriminator is a classifier with two classes- YES ($s=1$) and NO ($s=0$). 
Learning a discriminator (parameterized by $\phi$) involves taking a set of outputs of a model corresponding to a set of random inputs and assigning it to class $s=0$ if the model is non-Trojaned, and $s=1$ if it is Trojaned. 
We define $ \mathcal{\hat{D}}:=\cup_j \{(p(F_{N}(x_j;\theta''),T=1),0),(p(F_{T;\theta}(x_j),T=1),1) \}$, where $F_{N}(\cdot, \theta'')$ and $F_{T}(\cdot,\theta)$ are functions of non-Trojaned and Trojaned models parameterized by $\theta''$ and $\theta$, respectively. 
The discriminator minimizes a loss function derived from the cross-entropy loss $\mathcal{L}_{CE}$, given by: 
%
\begin{align}
    \mathcal{L}&=\frac{1}{2N'}\sum_{\substack{ (q,s)\in \mathcal{\hat{D}}\\ (q= p_{N},s=0) \:or  (q= p_{T},s=1) }} \mathcal{L}_{CE}(D(q;\phi),s)
\end{align}

The objective of the adversary, on the other hand, is to ensure that the backdoor in the Trojaned model remains undetectable (i.e., fool the discriminator). 
Consequently, she updates hyperparameters of the Trojaned model in a manner that will maximize the loss of the discriminator:
%
%
\begin{align}
    \max_\theta \min_{\phi} \mathcal{L}_{CE}(D(F(x;\theta);\phi),1)
\end{align}

Algorithm~\ref{alg:minmax} explains the min-max step in detail. 
In the $\min$ step, we first generate a set of arbitrary inputs (images, in our case that are generated from a $\mathcal{N}(\mu,\sigma)$ distribution) and provide them to both non-Trojaned and Trojaned models. 
The outputs from these models are used to train the discriminator. 
In the $\max$ step, we update parameters of the Trojaned model to maximize the loss of the discriminator by generating outputs that are similar to outputs of the non-Trojaned model for the arbitrarily generated inputs. 
In order to preserve accuracy of the model on any input (clean or triggered), we consider a set $\mathcal{D'}$ that contains both types of inputs, and minimize a cross-entropy loss (Lines 10-14). 

\begin{algorithm}
\caption{Min-Max Optimization for Discriminator}\label{alg:minmax}
\begin{algorithmic}[1]
\Require $F_{N}(., \theta''), F_{T}(.;\theta), \mu, \sigma, \mathcal{D'}, \alpha_1, \alpha_2, \alpha_3>0$

\For {$i=1:itr$}
\For {$i=1:N'$}
\State $img\leftarrow \mathcal{N}(\mu,\sigma)$
\State $\mathcal{\hat{D}} \leftarrow \mathcal{\hat{D}} \cup \{p((F_{N}(img;\theta''), T=1),0)\}$
\State $\mathcal{\hat{D}} \leftarrow \mathcal{\hat{D}} \cup \{(p(F_{T}(img;\theta), T=1),1)\}$
\EndFor
\State $L_1= \frac{1}{2N'}\sum_{(q,s)\in\mathcal{\hat{D}} } \frac{\partial}{\partial \phi} \mathcal{L}_{CE}(D(q;\phi),s)$
\State $\phi \gets \phi-\alpha_1 L_1$
\State $L_2=  \frac{1}{N'} \sum_{x_i\sim \mathcal{N}(\mu,\sigma)} \frac{\partial}{\partial \theta} \mathcal{L}_{CE}( D( p(F_T(x;\theta));\phi),1) $ 
\For {$(x^k,y^k) \in \mathcal{D'}$}
\State $z^k_{T} \gets F_{T}(x^k;\theta)$
\State $L_3=L_3+ \frac{\partial}{\partial \theta} \mathcal{L}_{CE} (y^k, p(z_{T}^k, T=1)) $
\EndFor
\State $\theta \gets \theta- \alpha_1 L_1 +\alpha_2 L_2 -\alpha_3 L_3$
\EndFor
\end{algorithmic}
\end{algorithm}

%% file: Experiments.tex
\section{Evaluation}\label{sec:evaluation}
%
%
%
\begin{table}[t]
    \centering
    \begin{tabular}{|c|c|c|c|}
    \hline
       Model &  \shortstack{Accuracy\\ on clean} &   \shortstack{backdoor \\success rate} &  \shortstack{Discriminator \\Detection Rate}\\ \hline
     Non-Trojaned model & {\pmb{$99.3\%$}} & - & $100\%$ \\ \hline
    Trojaned model &  $91.92\%$ & $81.06\%$ & $96.3\%$\\ \hline
    \shortstack{ Trojaned model with\\ Knowledge Distillation} & $98.74\%$ & $87.39\%$ & $99.40\%$\\ \hline
    Our approach & $90.21\%$ & \pmb{$96.82$} & \pmb{$0.0\%$}\\ \hline
    \end{tabular}
    \caption{Comparison of models: non-Trojaned, Trojaned trained with hard labels, Trojaned trained using only knowledge distillation, and Trojaned trained using knowledge distillation and max-min optimization (\textbf{ours}). Non-Trojaned models have the highest accuracy on inputs without a trigger (clean). Our approach results in the highest success rate for a multi-targeted backdoor attack, and completely bypasses a discriminator that aims to distinguish between outputs from a Trojaned and a non-Trojaned model ($0.0\%$ in last row).}
    \label{tab:accuracy}
\end{table}
This section introduces our simulation setup and then explains results of our empirical evaluations.
\subsection{Simulation Setup}
\begin{figure}
   \begin{tabular}{c c c c c}
       \includegraphics[scale=1.8]{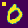} &  
       \includegraphics[scale=1.8]{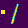}&
       \includegraphics[scale=1.8]{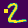}&
       \includegraphics[scale=1.8]{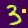}& 
       \includegraphics[scale=1.8]{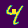}\\
        (0,1) & (1,2) & (2,3) & (3,4) & (4,5)  \\
        \includegraphics[scale=1.8]{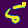} &  
       \includegraphics[scale=1.8]{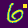}&
       \includegraphics[scale=1.8]{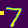}&
       \includegraphics[scale=1.8]{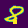}& 
       \includegraphics[scale=1.8]{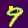}\\
       (5,6) & (6,7) & (7,8) & (8,9) & (9,0)  \\
   \end{tabular}
   \caption{The MNIST dataset that contains 10 classes, corresponding to the 10 digits. Each image is stamped with the trigger at a random location (yellow square). 
   The caption below each image shows (\emph{predicted label from non-Trojaned model}, \emph{predicted label from our Trojaned model}). The non-Trojaned model predicts the image labels correctly.}
    \label{fig:mnist}
\end{figure}

We use the MNIST dataset~\cite{lecun1998mnist} to evaluate Algorithms \ref{alg:KD} and \ref{alg:minmax}. 
This dataset contains $60000$ images of hand-written digits ($\{0,1,\cdots,,9\}$), of which $50k$ are used for training and $10k$ for testing, and each image is of size $28 \times 28$. 
A square of size $4\times4$ at an arbitrary location in the image is used as the trigger (shown in Figure~\ref{fig:mnist}). 

In order to learn a multi-target backdoor, we select a random subset of images from the training data that have been stamped with the trigger. 
Let $i$ denote the true class of the input that is stamped with the trigger, and $C$ denote the total number of classes ($C=10$ for MNIST). 
Then, these inputs are labeled according to $g(i):=(i+1) \mod C$. 
%

We use the recently proposed MTND defense~\cite{xu2021detecting} as a benchmark. 
MTND learns a discriminator that takes the output of a target model to return a `score'. 
If this score exceeds a pre-defined threshold, the model is identified as Trojaned, and is identified as non-Trojaned otherwise. 
%

The DNNs used to learn a classifier for the MNIST dataset consists of two convolutional layers, each constaining $5$ kernels, and channel sizes of $16$ and $32$ respectively. 
This is followed by maxpooling and fully connected layers of size $512$. 
For learning the discriminator, similar to~\cite{xu2021detecting}, we use a network with one (fully connected) hidden layer of size $20$. 
%
\begin{figure}[t]
    \centering
    \includegraphics[scale=0.5]{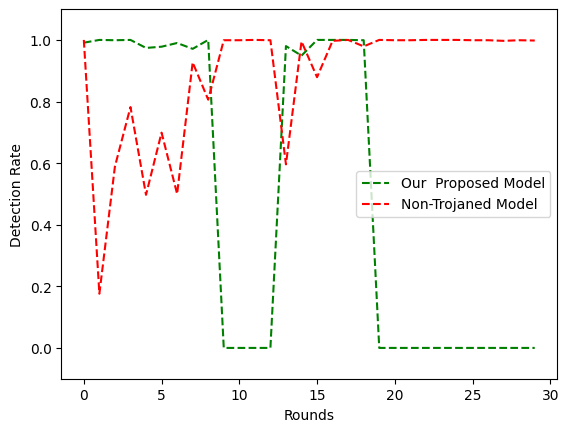}
    \caption{Detection rates of the discriminator for a non-Trojaned model (red) and our Trojaned model (green) during 30 rounds of Algorithm \ref{alg:minmax}. The discriminator is not able to optimize its hyper-parameters to detect both non-Trojaned model and our Trojaned model beyond $20$ rounds.}
    \label{fig:detectionrate}
\end{figure}
\subsection{Experiment Results} 
To demonstrate the limitations of existing defense mechanisms against backdoor attacks, we train the following models: 
(i) Trojaned, with multi-target backdoor using knowledge distillation and min-max optimization (\textbf{Our Trojaned Model}), 
(ii) Trojaned, with multi-target backdoor using only hard labels (\textbf{Traditional Trojaned Model}), and 
(iii) non-Trojaned, (\textbf{Non-Trojaned Model}). 
%

Table~\ref{tab:accuracy} indicates the accuracy of  these three models on clean inputs (i.e., images that are not stamped with a trigger), success rates of a multi-target  backdoor attack, and detection rates of a discriminator. 
The non-Trojaned model has the highest accuracy on clean inputs, since backdoor attacks decrease the accuracy of the models. 
Knowledge distillation is seen to improve the accuracy of Trojaned models on clean samples and success rates of a backdoor attack, but a Trojaned model trained using knowledge distillation alone can be detected by a discriminator with $99.4\%$ accuracy. 
When knowledge distillation is combined with min-max optimization, we see that the accuracy on clean inputs is reduced, but the success rate of the backdoor attack is higher. 
At the same time, the discriminator is not able to distinguish between outputs from a Trojaned and a non-Trojaned model ($0.0 \% $ in righmost column of last row). 
This demonstrates that the state-of-the-art MTND defense can be bypassed. 
%

Figure~\ref{fig:detectionrate} shows detection rates of the discriminator during different rounds of Algorithm \ref{alg:minmax}. 
The discriminator was not able to achieve a high detection rate on both Trojaned and non-Trojaned models when min-max optimization was deployed for more than 20 rounds.
\begin{figure}
    \centering
    \includegraphics[scale=0.5]{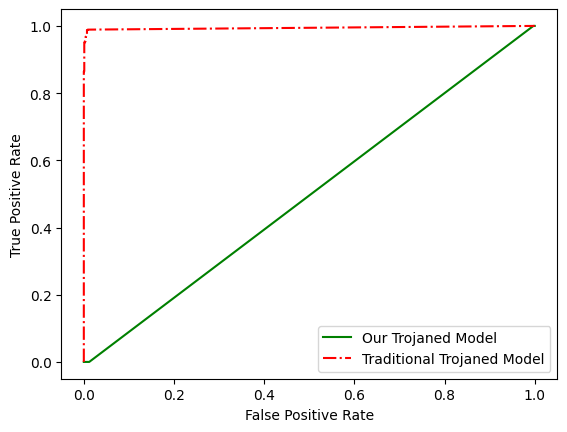}
    \caption{The ROC curve for different threshold values of the discriminator. There is no threshold that simultaneously returns low false positive and high true positive rates for detecting our Trojaned model. In comparison, for traditional Trojaned models,  a very small threshold  returns a low false positive rate and high true positive rate simultaneously.}
    \label{fig:auc}
\end{figure}

To evaluate the impact of the (pre-defined) threshold values associated with the discriminator on the detection rate, we plot and receiver operating characteristic (ROC) curve that compares the true positive rate (TPR) versus false positive rate (FPR) at different classification thresholds: 
\begin{equation*}
    TPR:= \frac{TP}{TP+FN}, \:\:\:  FPR:= \frac{TP}{TP+FN}
\end{equation*}
where $TP$ denotes the number of Trojaned models that are correctly identified as Trojaned and $FN$ is the number of non-Trojaned models that are incorrectly identified as Trojaned by the discriminator. 
%
The discriminator returns a score for each input. Inputs with a score exceeding a pre-defined threshold will be assigned to class $s=1$ (Trojaned). 
Figure~\ref{fig:auc} demonstrates that for any value of the threshold, the discriminator is not able to simultaneously return low $FPR$ and high $TPR$ for Trojaned models trained using our approach. 
However, for a traditional Trojaned model, the discriminator is able to simultaneously return a low $FPR$ and high $TPR$ for small threshold values. 

We also computed the area under the ROC curve (AUC) to measure the quality of the discriminator's predictions, independent of the chosen threshold value. 
The AUC is a number in the range $[0,1]$, and a model with an accuracy of $100 \%$ has $AUC=1$. 
We determined that our Trojaned model had $AUC = 0.495$, while the traditional Trojaned model had $AUC = 0.994$. 
These AUC values indicate that our two-step approach results in a discriminator performance in distinguishing between outputs from a Trojaned and non-Trojaned model that is as good as a `coin-toss' guess (i.e., selecting one of two possibilities, with probability $0.5$). 
%

%% file: discussion.tex
\section{Discussion}\label{sec:discussion}

In this section, we identify how our approach can be extended to domains where inputs might not be images, and highlight some open questions and challenges that are promising future research directions. 

\subsubsection{Extension to other domains} 
Our solution approach focused on the setup where inputs to a DNN was images, and we showed that the state-of-the-art MTND defense~\cite{xu2021detecting} could be bypassed. 
Recent research has demonstrated that DNNs designed for other tasks such as text classification and generation~\cite{dai2019backdoor}, code completion~\cite{schuster2021you}, and decision making in reinforcement learning and cyber-physical systems~\cite{panagiota2020trojdrl} are also vulnerable to backdoor attacks. 
We believe that our solution methodology can be applied to domains where the inputs and outputs of a DNN are continuous valued such as speech~\cite{zhai2021backdoor} and deep reinforcement learning~\cite{mnih2015human}. 
We will evaluate our two-step approach on backdoor attacks carried out on these types of systems, and examine the limitations of defenses against backdoor attacks in  settings where inputs to a DNN classifier are discrete-valued (e.g., text). 
%

\subsubsection{Provable Guarantees} 
The nested and inherently nonlinear structure of DNNs makes it challenging to explain the decision-making procedures for given input data~\cite{samek2017explainable}. 
This challenge also holds true when investigating the explainability of defense mechanisms and their limitations against backdoor attacks on DNNs. 
We believe that developing a principled approach that enables establishing provable guarantees on the  accuracy of certain classes of DNNs- e.g., when the activation function is a rectified linear unit (ReLU)~\cite{arora2018understanding} will be a promising step in this direction. 
%

\subsubsection{Building Better Defenses}
In this paper, the discriminator used outputs to an arbitrary set of image inputs to determine whether the model was Trojaned or not. 
An interesting question to answer is if the similarity between outputs from a Trojaned and non-Trojaned model can be characterized for any input that does not contain a trigger. 
Quantifying the change in the trigger/ trigger pattern that will result in a change in the decision of a Trojaned model compared to a non-Trojaned model is a possible solution approach. 
This will help learn and reason about the dynamics of the decision boundaries of DNN classifiers to enable building better defenses against backdoor attacks. 

%% file: conclusion.tex
\section{conclusion}\label{sec:conclusion}

This paper studied machine learning models that were vulnerable to backdoor attacks. 
Such a model is called a Trojaned model. 
We identified a limitation of a state-of-the-art defense mechanism that was
designed to protect against backdoor attacks. 
We proposed a new class of multi-target backdoor attacks in which a single trigger could result in misclassification to more than one target class. 
We then designed a two-step procedure that used knowledge distillation and min-max optimization to ensure that outputs from a Trojaned model were indistinguishable to  a non-Trojaned model. 
Through empirical evaluations, we demonstrated that our approach was able to completely bypass a 
state-of-the art defense mechanism, MTND. 
We demonstrated a reduction in detection accuracy of the discriminator of MTND from $>96\%$ without our method to $0\%$ when using our approach. 
We also discussed ways to extend our methodology to other classes of DNN models beyond those that use images, establish provable guarantees, and build better defenses.
%